\documentclass[aps,prl,twocolumn,superscriptaddress,showpacs]{revtex4}
\usepackage{bm}
\usepackage{epsf}
\usepackage{amssymb}
\usepackage{graphicx}
\usepackage{rotating}
\usepackage{epsfig}
\usepackage{psfrag}
\usepackage{amsmath}
\begin{document}
\def\G{{\cal G}}
\def\F{{\cal F}}
\def\bM{{\bm M}}
\def\bN{{\bm N}}
\def\bV{{\bm V}}
\def\bj{\bm{j}}
\def\bSig{{\bm \Sigma}}
\def\bLam{{\bm \Lambda}}
\def\bfeta{{\bf \eta}}
\def\bc{{\bf c}}
\def\ba{{\bf a}}
\def\d{{\bf d}}
\def\bP{{\bf P}}
\def\bK{{\bf K}}
\def\bk{{\bf k}}
\def\bkn{{\bf k}_{0}}
\def\bx{{\bf x}}
\def\bz{{\bf z}}
\def\bR{{\bf R}}
\def\br{{\bf r}}
\def\bu{{\bm u}}
\def\bq{{\bf q}}
\def\bp{{\bf p}}
\def\bG{{\bf G}}
\def\bQ{{\bf Q}}
\def\bs{{\bf s}}
\def\bA{{\mathbf A}}
\def\bv{{\bf v}}
\def\b0{{\bf 0}}
\def\la{\langle}
\def\ra{\rangle}
\def\Im{\mathrm {Im}\;}
\def\Re{\mathrm {Re}\;}
\def\beq{\begin{equation}}
\def\eeq{\end{equation}}
\def\bea{\begin{eqnarray}}
\def\eea{\end{eqnarray}}
\def\bdm{\begin{displaymath}}
\def\edm{\end{displaymath}}
\def\bnab{{\bm \nabla}}
\def\Tr{{\mathrm{Tr}}}
\def\bJ{{\bf J}}
\def\bU{{\bf U}}
\def\bPsi{{\bm \Psi}}
\def\mA {\mathrm{A}}

\title{Probing two-fluid hydrodynamics in a trapped Fermi superfluid at unitarity}

\author{Edward~Taylor}
\affiliation{Department of Physics, University of Toronto, Toronto, Ontario,
Canada, M5S 1A7}
\author{Hui~Hu}
\affiliation{Department of Physics, Renmin University of China, Beijing 100872, China}
\affiliation{ARC Centre of Excellence for Quantum-Atom Optics, Department of Physics, University of Queensland, Brisbane, Queensland 4072, Australia}
\author{Xia-Ji~Liu}
\affiliation{ARC Centre of Excellence for Quantum-Atom Optics, Department of Physics, University of Queensland, Brisbane, Queensland 4072, Australia}
\author{Allan~Griffin}
\affiliation{Department of Physics, University of Toronto, Toronto, Ontario,
Canada, M5S 1A7}
\date{\today}

\begin{abstract}
We develop a variational approach to calculate the density response function at finite temperatures of the lowest-lying two-fluid modes in a trapped two-component Fermi superfluid close to a Feshbach resonance. The out-of-phase oscillations, which are the analogue in trapped gases of second sound in uniform superfluids, have so far not been observed in cold-atom experiments.  At unitarity, we show that these modes are observable at finite temperatures via two-photon Bragg scattering, whose spectrum is related to the imaginary part of density response function.  This provides direct evidence for superfluidity and a promising way to test microscopic results for thermodynamics at unitarity.
\end{abstract}
\maketitle

There is growing experimental evidence~\cite{Thomas05,Grimm07,Thomas07} that trapped Fermi gases close to a Feshbach resonance are in the collisional hydrodynamic regime.  It has been argued that the finite temperature dynamics in this region should be described by the Landau two-fluid hydrodynamic equations~\cite{TaylorPRA05}.  Landau's two-fluid hydrodynamics~\cite{Landau41, Khalatnikov} describes the coupled dynamics of the superfluid and normal fluid components when collisions produce a state of local thermodynamic equilibrium.  A Feshbach resonance can be used to increase the magnitude of the \textit{s}-wave scattering length between Fermi atoms in two different hyperfine states. Thus, close to the unitarity limit where the scattering length diverges, the Landau two-fluid description of finite temperature collisional dynamics of trapped Fermi superfluids is expected to be valid. 
In a recent paper~\cite{TaylorPRA08}, we have used a variational approach to calculate the two-fluid dipole and breathing modes.  In this Letter, we extend this variational theory to calculate the density response function given by Landau's two-fluid hydrodynamics in a trapped Fermi gas at unitarity.  We propose that two-photon Bragg scattering~\cite{Pitaevskiibook} (which has been used to study the collisionless region~\cite{KatzPRL2004,DavidsonRMP05,KetterlePRL99}) is an ideal tool to probe Landau two-fluid behaviour.  

The most spectacular prediction of Landau's two-fluid hydrodynamics is the existence of second-sound, an out-of-phase oscillation of the normal and superfluid components. Due to weak interactions, this region has never been probed in Bose-condensed gases. For a strongly interacting Fermi gas near unitarity, one can show that the recent studies of breathing modes have probed the in-phase two-fluid mode~\cite{Thomas04,Grimm04,Thomas05} since \emph{only} this mode is excited by modulating the harmonic confining potential.  We have shown in Ref.~\cite{TaylorPRA08} that the frequency of this in-phase mode is temperature independent at unitarity.  This result explains why these experiments are well explained by the oscillations of a pure superfluid at $T=0$~\cite{Stringari04EPL, Heiselberg04,Tosi04}.  The out-of-phase modes predicted by two-fluid hydrodynamics in trapped Fermi gases at finite temperatures are much more interesting, being the analogue of second sound.

The Bragg scattering cross section is proportional to the imaginary part Im$\chi_{\rho\rho}(\bq,\Omega)$ of the density response function, where $\bq$ and $\Omega$ are the wave vector and frequency difference between two laser beams.  In experiments done so far on trapped Bose gases, Bragg scattering has been used both in the large $q$ ($\equiv|\bq|$) region ($qR_{TF}\gg 2\pi$, $R_{TF}$ being the smallest Thomas-Fermi radius of the condensate) where the phonon excitations characteristic of uniform systems are probed~\cite{KetterlePRL99}, and in the small $q$ region such that $qR_{TF} \sim O(1)$~\cite{KatzPRL2004}. In the latter region, the wavelength $\lambda=2\pi/q$ of the Bragg beams is comparable to the size of the condensate, and therefore the lowest energy normal modes can be excited.

The variational approach we develop (see also Ref.~\cite{TaylorPRA05}) is well suited for dealing with the two-fluid hydrodynamics at finite temperatures in a spatially non-uniform trapped Fermi gas.  We concentrate on the BCS-BEC crossover at unitarity since this is perhaps the most interesting region.  However, our formalism is applicable away from unitarity as long as the collisions are strong enough to produce local equilibrium.  
Our new results show that the density response spectrum exhibits well defined resonances corresponding to the out-of-phase oscillations at temperatures of the order $0.5T_c$ in the superfluid phase.  The frequencies and spectral weights of the \textit{out-of-phase} modes are very dependent on the temperature.  The weight vanishes both at $T=0$ and above $T_c$.  The \textit{in-phase} hydrodynamic modes frequencies are independent of temperature~\cite{TaylorPRA08} but the spectral weights change with $T$.

\textit{Out-of-phase mode frequencies}. --- Before discussing the density response function, we need to review the variational theory~\cite{TaylorPRA05,TaylorPRA08} for the solutions of the Landau two-fluid equations.  This is based on an action $S^{(2)}_0$ that describes fluctuations in the displacement fields $\bu_s,\bu_n$ defined by $\bv_s \equiv \dot{\bu}_s$ and $\bv_n \equiv \dot{\bu}_n$, where $\bv_s$ and $\bv_n$ are the superfluid and normal fluid velocities, respectively. 
The stationary point of the action gives the solutions of the linearized Landau two-fluid differential equations and hence, the two-fluid hydrodynamic mode frequencies.  

Following Refs.~\cite{TaylorPRA05,TaylorPRA08}, we make a physically motivated ansatz for the displacement fields for the low energy hydrodynamic modes in trapped gases based on the known exact solutions for a pure superfluid ($T=0$) and a normal fluid ($T>T_c$).  In these two limits, the exact solutions are of the same form.  For the \textit{dipole} mode, characterized by displacements of the centre-of-masses of the normal and superfluid components along one of the axes of the harmonic trap (we choose the $z$ axis), we use $\bu_{s,n}(\br,t)=\sum_{\omega}e^{-i\omega t}a_{s,n}(\omega)\hat{\bz}$, where $\hat{\bz}$ is the unit vector along the $z$ axis.  For the \textit{breathing} mode in an isotropic trap, we use the scaling ansatz $\bu_{s,n}(\br,t)=\sum_{\omega}e^{-i\omega t}a_{s,n}(\omega)\br$.  For both these modes, the action can be written in terms of the variational parameters $a_s$ and $a_n$ as 
\bea S^{(2)}_0 = \frac{1}{2}\sum_{\omega}(a^{*}_s,a^{*}_n)\bA(\omega)(a_s,a_n)^T, \label{s2} \eea
where we have defined the  $2\times 2$ matrix $\bA$ as 
\bea \bA(\omega) \equiv\left [
\begin{array}{cc} \! M_{s}\omega^2\!-\!k_s & -k_{sn} \\ -k_{sn} &\! M_{n}\omega^2 \!-\!k_n
\end{array}\! \right ], \label{bA}\eea and the superscript ``$T$" denotes the transpose of the two component spinor.  
The superfluid and normal fluid mass moments, $M_s$ and $M_n$, as well as the ``spring constants" $k_s$, $k_n$, and $k_{sn}$, depend on the choice of ansatz for different normal modes~\cite{TaylorPRA05,TaylorPRA08}. These are given below for the dipole and breathing modes.  The variational solutions (denoted by the superscript ``L") of the Landau two-fluid equations are given by
\bea \frac{\partial S^{(2)}_0}{\partial a^{*}_{s}}\Big|_{\!a_{s}=a^L_{s}} = \frac{\partial S^{(2)}_0}{\partial a^{*}_{n}}\Big|_{\!a_{n}=a^L_{n}}=0, \label{EL}\eea
which leads to $\bA(a^L_{s},a^L_{n})^T=0$. 
The frequencies of the in-phase ($\omega_1$) and out-of-phase ($\omega_2$) modes are thus given by the solution of $\mathrm{det}\bA(\omega)\! =\! 0$.  
As shown in Ref.~\cite{TaylorPRA08}, for both the dipole and breathing modes at unitarity, the in-phase and out-of-phase modes are described by the displacements $a_s = a_n$ and $M_sa_s + M_na_n = 0$, respectively.  

For the dipole mode, one finds~\cite{TaylorPRA08} $M^{D}_s\equiv \int d\br\;\rho_{s0}(\br)$, $M^{D}_{n}\equiv \int d\br\;\rho_{n0}(\br)$, $k^{D}_{sn}=M^{D}_s\omega^2_0-k^{D}_{s}$, and $k^{D}_n=(M^{D}_n-M^D_s)\omega^2_0+k^{D}_{s}$, where we have defined
\bea 
k^{D}_{s}\equiv\int d\br\;\left(\frac{\partial
\mu}{\partial\rho}\right)_{\!s}\frac{\partial \rho_{s0}}{\partial
z}\!\frac{\partial \rho_{s0}}{\partial z}. \label{ksnmt}\eea
Here, $\rho_{s0}$ and $\rho_{n0} \equiv \rho_{0} - \rho_{s0}$ are the equilibrium superfluid and normal fluid densities, respectively.  Using these results, the in-phase dipole mode is just the generalized Kohn mode with frequency $\omega_{1D}=\omega_0$ equal to the trap frequency at all temperatures.  The frequency of the out-of-phase dipole oscillation is  
\bea \omega^{2}_{2D}= \frac{k^{D}_{s}}{M^{D}_r} - \frac{M^D_s}{M^D_n}\omega^2_0, \label{mode2dm} \eea
where $M^{D}_r \equiv M^{D}_sM^{D}_n/(M^{D}_s+M^{D}_n)$ is the dipole mode reduced mass.

At unitarity, the universal nature of the thermodynamics~\cite{Ho04} leads to great simplifications in the expression for the breathing mode frequencies.  For an isotropic trap, we find~\cite{TaylorPRA08}, $M^{B}_s\equiv \int d\br\;\rho_{s0}(\br)r^2$, $M^{B}_{n}\equiv \int d\br\;\rho_{n0}(\br)r^2$, $k^{B}_{sn}=4M^{B}_s\omega^2_0-k^{B}_s$, and $k^{B}_{n}=4(M^{B}_n-M^{B}_s)\omega^2_0+k^{B}_s$, where we have defined
\bea k^{B}_s &\equiv&\int d\br\;\left(\frac{\partial\mu}{\partial \rho}\right)_{\!s}\left\{\bnab\cdot\left[\br \rho_{s0}(\br)\right]\right\}^2.\label{ksiso}\eea 
The frequency of the in-phase mode at unitarity is independent of temperature, given by $\omega_{1B}\! =\! 2\omega_0$~\cite{Castin}.  The frequency of the out-of-phase mode is 
\bea \omega^2_{2B} = \frac{k^{B}_{s}}{M^{B}_r} - 4\frac{M^{B}_{s}}{M^{B}_n}\omega^2_0,\label{mode2bm}\eea
with the breathing mode reduced mass $M^{B}_r \equiv M^{B}_sM^{B}_n/(M^{B}_s + M^{B}_n)$.   The frequencies of both this out-of-phase breathing mode and the dipole mode [Eq.~(\ref{mode2bm})] are strongly dependent on temperature (see Figs.~3 and 4 of Ref.~\cite{TaylorPRA08}).

\textit{Density response function}. --- The two-photon Bragg scattering spectrum is well known to be directly related to the imaginary part of the density response function $\chi_{\rho\rho}(\bq,\Omega)$~\cite{KetterlePRL99,Tozzo}.  This in turn is related to the antisymmetric part of the dynamic structure factor $S(\bq,\Omega)$,\bea\!\!\!\!\!-\frac{\mathrm{Im}\chi_{\rho\rho}(\bq,\Omega)}{2\pi m^2 N} =\frac{1}{2}\left[S(\bq,\Omega)\! -\! S(\bq,-\Omega)\right]\equiv
S_A(\bq,\Omega),\label{SA} \eea where $N$ is the number of Fermi atoms.   
This formalism is valid in both the collisionless and hydrodynamic regions.  In this Letter, we concentrate on the contribution to $S_A$ from the hydrodynamic dipole and breathing modes in trapped gases.

To find the density response function $\chi_{\rho\rho}(\bq,\Omega)$, we consider an external potential of Bragg beams which excites modes of frequency $\Omega$ and wavevector $\bq$: $V_{\mathrm{pert}}(\br,t)=V_{\bq,\Omega}e^{i(\bq\cdot\br-\Omega t)}$. Assuming this is a weak perturbation, there will be a density response of the form 
\bea \delta\rho(\bq,\Omega) &=& \chi_{\rho\rho}(\bq,\Omega)V_{\bq,\Omega}, \label{response}\eea
which defines the linear density response function $\chi_{\rho\rho}$.  Our task is to find the structure of $\chi_{\rho\rho}(\bq,\Omega)$ imposed by the Landau two-fluid equations in the presence of both $V_{\mathrm{pert}}$ and the trap potential.

The Bragg perturbing potential generates an extra term $S^{(2)}_{\mathrm{pert}} = -\int d\br dt [V_{\mathrm{pert}}(\br,t)\delta\rho^{+}(\br,t) + \mathrm{h.c.}]$ in the action, with $\delta \rho(\br,t) = -\mbox{\boldmath$\nabla$}\cdot\left[\rho_{s0}(\br)\bu_s(\br,t) + \rho_{n0}(\br)\bu_n(\br,t)\right]$ being the density fluctuation operator~\cite{TaylorPRA05}. Thus,
\bea S^{(2)}_{\mathrm{pert}}=V_{\bq,\Omega}\left[F^{*}_{s}(\bq)a^{*}_{s}(\Omega) + F^{*}_{n}(\bq)a^{*}_{n}(\Omega)\right] + \mathrm{h.c.}, \label{deltaS3b} \eea
with the weighting factors \bea F^D_{s,n}(\bq) &\equiv& -iq_z \int d\br e^{i\bq\cdot\br}\rho_{s0,n0}(\br)\nonumber\\ F^B_{s,n}(\bq) &\equiv& -i\int d\br e^{i\bq\cdot\br}(\bq\cdot\br)\rho_{s0,n0}(\br)\label{Fsn}\eea for the dipole (D) and isotropic breathing modes (B).  
The variational solutions of the Landau two-fluid equations are now given by Eq.~(\ref{EL}), using the action $S^{(2)}_0 \!+\! S^{(2)}_{\mathrm{pert}}$.  It is straightforward to show that this gives $(a^L_{s},a^L_{n})^T=-\bA^{-1}\left[F^{*}_{s}(\bq), F^{*}_{n}(\bq)\right]^{T}V_{\bq,\Omega}$ for the variational parameters describing the two-fluid modes.
Hence, the density response of the Landau two-fluid equations, $\delta\rho_L(\bq,\Omega)=-\left[F_{s}(\bq)a_{s}(\Omega) + F_{n}(\bq)a_{n}(\Omega)\right]$, is given by Eq.~(\ref{response}) but now we have an \textit{explicit} expression for the density response function in the Landau two-fluid region,
\bea \chi^L_{\rho\rho}(\bq,\Omega)\!\! &=&\!\! \sum\nolimits_{\alpha\beta} F^{*}_{\alpha}[\bA^{-1}(\Omega)]_{\alpha\beta}F_{\beta}, \label{chiqw} \eea 
where $[\bA^{-1}(\Omega)]_{\alpha\beta}$ ($\alpha,\beta = s,n$) denotes the $\alpha\beta$ element of the inverse of the matrix $\bA(\Omega)$ defined in Eq.~(\ref{bA}).

Using Eq.~(\ref{bA}) in Eq.~(\ref{chiqw}), the density fluctuation spectrum given by Eq.~(\ref{SA}) becomes (for $\Omega>0$), 
\bea S^{H}_{\!A}(\bq,\Omega)\!=\!\gamma_{\bq}\!\!\left[\frac{Z_1(\bq)}{2\omega_1}\delta(\Omega\!-\!\omega_1)\!+\!\frac{Z_2(\bq)}{2\omega_2}\delta(\Omega\!-\!\omega_2)\right],
\label{Imchi}\eea
where \bea \gamma_{\bq}\equiv\frac{1}{2m^2N}\left[\frac{\left|F_{s}(\bq)\right|^2}{M_s}+\frac{\left|F_{n}(\bq)\right|^2}{M_n}\right],\label{gammaq}\eea and the relative weight of the in-phase mode $\omega_1$ is given by
\bea Z_1(\bq) \equiv \frac{\left|F_{s}(\bq) + F_{n}(\bq)\right|^2M_r}{\left|F_{s}(\bq)\right|^2M_n + \left|F_{n}(\bq)\right|^2M_s}.\label{Z1}\eea
The functions $F_{n,s}(\bq)$ are defined in Eq.~(\ref{Fsn}).  The out-of-phase mode $\omega_2$ in Eq.~(\ref{Imchi}) has relative weight $Z_2(\bq) \equiv 1 - Z_1(\bq)$.  
Equations~(\ref{Imchi}) and (\ref{Z1}) are valid for both the dipole (with $F_{s,n} = F^D_{s,n}$ and $M_{s,n} = M^D_{s,n}$) and breathing ($F_{s,n} = F^B_{s,n}$ and $M_{s,n} = M^B_{s,n}$) modes.  Equation~(\ref{Imchi}), describing the density response function predicted by Landau two-fluid hydrodynamics, is the key result of this Letter.  It is the natural generalization of the well-known expression~\cite{HohenbergPRL} for a \textit{uniform} superfluid, to be discussed shortly.  

\begin{figure}
\begin{center}
\epsfig{file=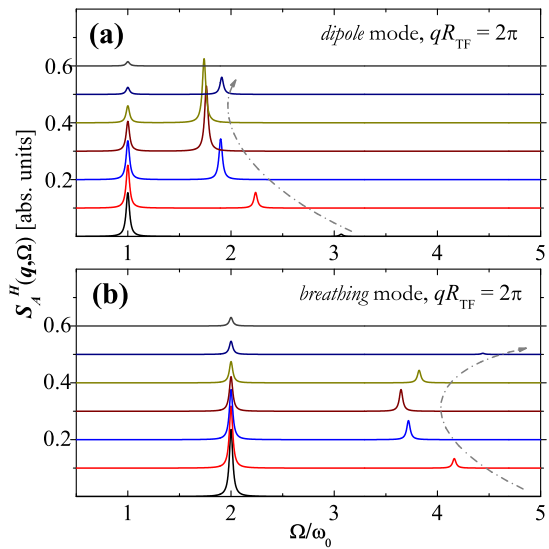, angle=0,width=0.40\textwidth}
\caption{(Color online) Temperature dependence of the density response spectrum as a function of the frequency $\Omega$ at unitarity due to (a) dipole and (b) breathing modes, for an isotropic trap with frequency $\omega_0$. The momentum transferred $q = 2\pi/R_{TF}$ is given in terms of the Thomas-Fermi radius of a ideal trapped Fermi gas.  From bottom to top, the temperature $T$ increases from 0.04$T_F$ to 0.28$T_F$ with a step 0.04$T_F$, where $T_F$ is the Fermi temperature of an ideal trapped Fermi gas. The top curve is above the superfluid transition temperature $T_c\simeq 0.27T_F$.}
\label{tkappa}
\end{center}
\end{figure}

\begin{figure}
\begin{center}
\epsfig{file=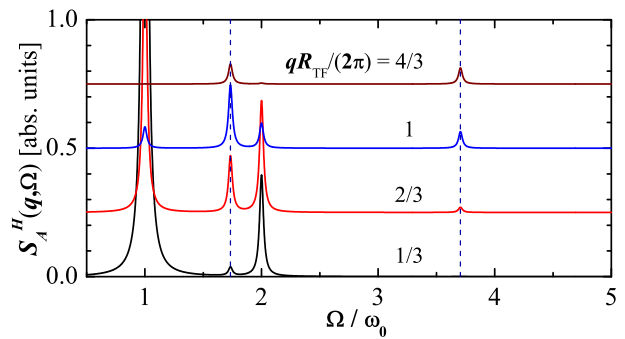, angle=0,width=0.45\textwidth}
\caption{(Color online) The spectral function $S_A(\bq,\omega)$ with increasing Bragg momentum $q$, for $T=0.7T_c$.}
\label{qkappa2}
\end{center}
\end{figure}

Numerical results for $S^{H}_A(\bq,\Omega)$ for the dipole and breathing modes for several different temperatures are shown in Fig.~\ref{tkappa}. 
We use the thermodynamic functions and superfluid density for a uniform superfluid given in Ref.~\cite{TaylorPRA08} and use a local density calculation approximation to calculate these quantities in a trapped gas.  
The results are quite dependent on the momentum transfer $q = 2\pi/\lambda$ for the Bragg pulse.  This is shown in Fig.~\ref{qkappa2}, which compares $S^{H}_A(\bq,\Omega)$ for several different values of $q$, at the temperature $T=0.7T_c$.  
In Figs.~\ref{tkappa} and \ref{qkappa2}, the width of the resonances is taken to be $0.02\omega_0$.  In experiments, the width will go as $2\pi/\Delta t$, where $\Delta t$ is the Bragg pulse duration (see, for example, Ref.~\cite{Tozzo}).  
\begin{figure}
\begin{center}
\epsfig{file=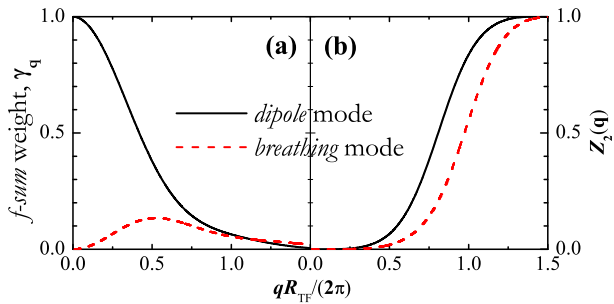, angle=0,width=0.45\textwidth}
\caption{Momentum dependence of (a) the \textit{f}-sum weight $\gamma_{\bq}$ (in units of $q^2/4m$) and (b) weight $Z_2$ of the lowest energy out-of-phase dipole and breathing modes at $T=0.18T_F\simeq0.7T_c$.}
\label{qkappa3}
\end{center}
\end{figure}

As $T\!\!\rightarrow\!\!0$, $M_r\!\! \rightarrow\! \!M_n$, and the weights of both the in-phase dipole and breathing modes goes to unity, $Z_1\!\!\rightarrow\!\! F^2_s(\bq)M_r/F^2_s(\bq)M_n\! \!=\! \!1$.  Thus, $S^{H}_A(\bq,\omega)$ only has resonances at the frequencies of the in-phase modes at $T=0$.  As the temperature increases, however, the out-of-phase modes acquire \textit{finite} weight and $S^{H}_A(\bq,\Omega)$ gains additional resonances at these frequencies.  As $T\rightarrow T_c$, $M_r \rightarrow M_s$, and $Z_1 \rightarrow F^2_n(\bq)M_r/F^2_n(\bq)M_s = 1$. As expected, above $T_c$, only the in-phase modes are present. 

In understanding $S_A(\bq,\Omega)$ in both uniform and trapped gases, the well-known $f$-sum rule~\cite{NozieresPines2,Pitaevskiibook} is very useful.  Quite generally, $S_A(\bq,\Omega)$ satisfies the condition
\bea \int^{\infty}_0 d\Omega \;\Omega S_A(\bq,\Omega) = q^2/4m.\label{fsum}\eea
The spectral density $S_A$ has contributions from all hydrodynamic modes, as well from high energy excitations not described by the two-fluid equations.  For a trapped gas, the two-fluid density response in Eq.~(\ref{Imchi}) gives
\bea  \int^{\infty}_0 d\Omega \;\Omega S^H_A(\bq,\Omega) = \gamma_{\bq}(T),\eea
which shows that $\gamma_{\bq}$ in Eq.~(\ref{gammaq}) describes the relative contribution of a hydrodynamic mode to the $f$-sum rule for a given value of $\bq$.   

In Fig.~\ref{qkappa3}, we plot the dependence on $\bq$ of the weight factor $\gamma_{\bq}$ in Eqs.~(\ref{Imchi}) and (\ref{Z1}) of the dipole and breathing hydrodynamic modes as well as the weight $Z_2(\bq)$ of the out-of-phase modes.  These results are for $T=0.7T_c$ and are quite dependent on temperature (as can be seen in Fig.~\ref{tkappa}).  The optimal momentum to observe the out-of-phase mode is seen to be about $2\pi/R_{TF}$, where the overall weight $\gamma_{\bq}Z_2(\bq)$ reaches a maximum.  As shown in Fig.~\ref{qkappa3}(a), $\gamma_{\bq}$ is significantly reduced from the maximum value of $q^2/4m$ when $q \sim 2\pi/R_{TF}$.  This means that at larger values of $q$, the spectral weight in $S_A(\bq,\Omega)$ is shifting to other low-energy hydrodynamic modes (e.g., quadrupole) as well as high-energy excitations not described by two-fluid hydrodynamics.  The successful observation of the out-of-phase modes using Bragg scattering will require trying to find the optimal values of $q$ and $T$.   
  
\begin{figure}
\begin{center}
\epsfig{file=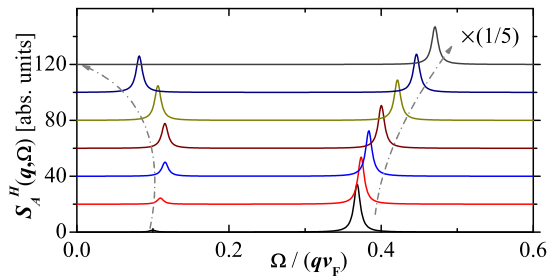, angle=0,width=0.40\textwidth}
\caption{(Color online) Temperature dependence of the density response spectrum in a uniform gas as a function of the frequency $\Omega$. The offset spectra go from $T=0.4T_c$ to $T=T_c$ in steps of $0.1T_c$.  The spectral weight of second sound (lower branch) is very small below $0.4T_c$.  The spectral weight of first sound is everywhere much larger than second sound and has been reduced by a factor of 5 in this plot.}
\label{tkappauniform}
\end{center}
\end{figure}

Using a plane-wave ansatz for the displacement fields, one can verify that Eq.~(\ref{chiqw}) gives the usual two-fluid density response function for a \textit{uniform} superfluid~\cite{HohenbergPRL}. In Fig.~\ref{tkappauniform}, we show $S^{H}_A(\bq,\Omega)$ for a uniform gas at unitarity, based on the \textit{same} thermodynamic functions~\cite{TaylorPRA08} used to compute the density fluctuation spectrum for a trapped gas (using a local density approximation) shown in Figs.~\ref{tkappa}-\ref{qkappa3}.  For a uniform gas, the poles of $S^{H}_A(\bq,\Omega)$ correspond to first and second sound ($\omega_i = c_i q$), the analogue of the in-phase and out-of-phase modes in a trap which we have been discussing. In a uniform superfluid, the expression for $S^H_A(\bq,\Omega)$ describing first and second sound modes saturates the $f$-sum rule in Eq.~(\ref{fsum}), namely
\bea \int^{\infty}_0 d\Omega \;\Omega S^H_A(\bq,\Omega) = q^2/4m.\label{fsumuni}\eea However, the nature of both the in-phase and out-of-phase modes is quite different in traps and there is no simple comparison.  
In contrast to the results for a trapped gas shown in Fig.~\ref{tkappa}, the second sound frequency goes to zero at $T_c$ in a uniform gas.  The lowest temperature shown in Fig.~\ref{tkappauniform} is $0.4T_c$, since the weight of second sound is vanishingly small at lower temperatures.  However, we note that the second sound velocity $c_2$ increases to the $T=0$ value $c_1/\sqrt{3} \simeq 0.21v_F$, where $c_1 \simeq 0.37v_F$ is equivalent to the Bogoliubov phonon velocity at $T=0$ (see, for instance, Ref.~\cite{Heiselberg05}).

\textit{Conclusions}. --- We have proposed that two-photon Bragg scattering is an ideal tool to probe the Landau two-fluid superfluid hydrodynamics in a trapped strongly interacting Fermi gas near unitarity.  To this end, we have presented explicit predictions for the density response spectrum which can be measured with available experimental techniques~\cite{KatzPRL2004,DavidsonRMP05,KetterlePRL99}. Previous studies for uniform~\cite{Ho04,Heiselberg05} and trapped~\cite{Levin07} gases did not calculate the spectral weights of the hydrodynamic modes in the dynamic structure factor.  The observation of second sound-like out-of-phase modes at finite temperatures is one of the challenges in current research on ultracold atomic gases. 

E.T. and A.G. thank J.~Thywissen for useful discussions on Bragg scattering. They are supported by NSERC of Canada. H.H. and X.-J.L. are supported by the National Fundamental Research Program of China Grant Nos. 2006CB921404 and 2006CB921306, and the Australian Research Council Center of Excellence.


\begin{thebibliography}{99}
\bibitem{Thomas05} J.~Kinast \textit{et al.}, Phys. Rev. Lett. \textbf{94}, 170404 (2005).
\bibitem{Grimm07} M.~J. Wright \textit{et al.}, Phys. Rev. Lett. \textbf{99}, 150403 (2007).
\bibitem{Thomas07} A.~Turlapov \textit{et al.}, e-print: arXiv:0707.2574.
\bibitem{TaylorPRA05} E.~Taylor and A.~Griffin, Phys. Rev. A \textbf{72}, 053630 (2005). 
\bibitem{Landau41} L.~D.~Landau, J. Phys. U.S.S.R. \textbf{5}, 71 (1941).
\bibitem{Khalatnikov} I.~M.~Khalatnikov, \textit{An Introduction to the Theory of Superfluidity} (W.~A.~Benjamin, New York, 1965).
\bibitem{TaylorPRA08} E.~Taylor, H.~Hu, X.-J.~Liu, and A.~Griffin, e-print: arXiv:0711.0561.
\bibitem{Pitaevskiibook} S.~Stringari and L.~Pitaevskii, \textit{Bose-Einstein Condensation} (Oxford University Press, Oxford, 2003), p.~197. 
\bibitem{KatzPRL2004} N.~Katz \textit{et al.}, Phys. Rev. Lett. \textbf{93}, 220403 (2004).
\bibitem{DavidsonRMP05}  R.~Ozeri \textit{et al.}, Rev. Mod. Phys. \textbf{77}, 187 (2005).
\bibitem{KetterlePRL99} D.~M.~Stamper-Kurn \textit{et al.}, Phys. Rev. Lett \textbf{83}, 2876 (1999).
\bibitem{Thomas04} J.~Kinast \textit{et al.}, Phys. Rev. Lett. \textbf{92}, 150402 (2004).
\bibitem{Grimm04} M.~Bartenstein \textit{et al.}, Phys. Rev. Lett. \textbf{92}, 203201 (2004).
\bibitem{Stringari04EPL} S.~Stringari, Europhys. Lett. \textbf{65}, 749 (2004).
\bibitem{Heiselberg04} H.~Heiselberg, Phys. Rev. Lett. \textbf{93}, 040402 (2004).
\bibitem{Tosi04} H.~Hu \textit{et al.}, Phys. Rev. Lett. \textbf{93}, 190403 (2004). 
\bibitem{Ho04} T.-L.~Ho, Phys. Rev. Lett. \textbf{92}, 090402 (2004).
\bibitem{Castin} Y.~Castin, Comptes Rendus Physique \textbf{5}, 407 (2004).  
\bibitem{Tozzo} C.~Tozzo and F.~Dalfovo, New. J. Phys. \textbf{5}, 54 (2003). 
\bibitem{HohenbergPRL} P.~C.~Hohenberg and P.~C.~Martin, Phys. Rev. Lett. \textbf{12}, 69 (1964).
\bibitem{NozieresPines2} P.~Nozi\`{e}res and D.~Pines, \textit{The Theory of Quantum Liquids, vol. II} (Addison-wesley, Redwood City, 1989). 
\bibitem{Heiselberg05} H.~Heiselberg, Phys. Rev. A \textbf{73}, 013607 (2006).
\bibitem{Levin07} Y.~He \textit{et al.}, Phys. Rev. A \textbf{76}, 051602(R) (2007). 
\end{thebibliography}
\end{document}